# The emergence of dynamic networks from many coupled polar oscillators. A model for Artificial Life.

**Alessandro Scirè\* · Valerio Annovazzi-Lodi**

**Abstract**
This work concerns a many-body deterministic model that displays life-like properties as emergence, complexity, self-organization, spontaneous compartmentalization, and self-regulation. The model portraits the dynamics of an ensemble of locally coupled polar phase oscillators, moving in a two-dimensional space, that in certain conditions exhibit emergent superstructures. Those superstructures are self-organized dynamic networks, resulting from a synchronization process of many units, over length scales much greater than the interaction length. Such networks compartmentalize the two-dimensional space with no a priori constraints, due to the formation of porous transport walls, and represent a highly complex and novel non-linear behavior. The analysis is numerically carried out as a function of a control parameter showing distinct regimes: static, stable dynamic networks, intermittency, and chaos. A statistical analysis is drawn to determine the control parameter ranges for the various behaviors to appear. The model and the results shown in this work are expected to contribute to the field of artificial life.

\* Corresponding author
Email: alessandro.scire@unipv.it (A.S.)

## 1 Introduction

*Artificial Life* (ALife) is an interdisciplinary research topic (Langton, 1997; Adami, 1998; Dorin, 2014), that brings together scientists, philosophers and artists. Three interplaying branches of artificial life are commonly devised: "Soft" artificial life creates numerical simulations that exhibit life-like behavior, "hard" artificial life produces hardware implementations of life-like systems, and "wet" artificial life synthesizes living systems from biochemical products (Rasmussen et al., 2003, 2008).

Alessandro Scirè
Dipartimento di Ingegneria Industriale e dell'Informazione, Università di Pavia, I - 27100, Pavia, Italy.
E-mail: alessandro.scire@unipv.it

Valerio Annovazzi-Lodi
Dipartimento di Ingegneria Industriale e dell'Informazione, Università di Pavia, I - 27100, Pavia, Italy.
E-mail: valerio.annovazzi@unipv.it

Several fields are involved, such as complexity (Bar-Yam, 1997; Mitchell, 2009), natural computing (Castro, 2006), evolutionary computation (Baeck et al., 1997; Coello et al., 2007), language evolution (Christiansen and Kirby, 2003; Cangelosi and Parisi, 2002), theoretical biology (Waddington, 1968), evolutionary biology (Smith et al., 1995), philosophy (Boden, 1996), cognitive science (Clark, 1997; Bedau, 2003; Couzin, 2009), robotics (Mataric and Cliff, 1996), artificial intelligence (Steels and Brooks, 1995), behavior-based systems (Maes, 1993; Webb, 2000), game theory (Sigmund, 1993), network theory (Newman, 2003; Newman et al., 2006), and synthetic biology (Benner and Sismour, 2005) among others.

The term Artificial Life (Alife) was introduced (Langton, 1989) as "life made by man rather than by nature", meaning artificial systems that exhibit life-like properties. More recently (Bedau, 2007) defined artificial life as an interdisciplinary research concerning life and life-like processes, that emphasizes the inherent/organizational rather than the structural/material properties of living systems, and that aims at comprehending living systems by creating simple forms of them. Meaning to create simple artificial systems that display some specific life-like properties such as compartmentalization, homeostasis, the ability to reproduce, growth and development, adaptation, and evolution, among many others. Compartmentalization, in particular, is considered of fundamental importance for life. Indeed, primitive compartments provided a mechanism by which chemical systems underwent speciation. *"It is indeed unlikely that life started in […] conditions of extreme dilution of a few molecules in the prebiotic ocean, and some form of compartmentalization might be considered to explain how the necessary local metabolite concentration was achieved."* (Luisi, 2014).

However, many of the specific life-like properties are paraphrases of two generic processes, namely *emergence* and *self-organization*, which are indeed both believed to lay at the root of abiogenesis (Luisi, 2006). Emergence refers to a collective behavior that is *more than the sum of the parts*, what parts of a system do together that they would not do alone, whereas self-organization means a process where order arises solely from local interactions, with no *Deus ex machina* intervention. Soft ALife has been linked to emergence and self-organization in many subdomains. Cellular automata, a popular form of soft ALife, are illustrative examples of self-organizing systems. Without a global controller involved, cellular automata self-organize their state configurations in many ways (Wolfram, 2002). Further examples, among others, are Partial Differential Equations (PDEs) (Cross and Hohenberg, 1993) and self-propelled agents (Krolikowski, 2016), that show a wide range of self-organizing dynamics and emergent properties.

Concerning dynamical systems and PDEs, self-organization and emergence are expressed by spontaneous pattern formation and spatio-temporal coherence. This means the spontaneous synchronization of the spatiotemporal dynamics of many units, where "spontaneous" means emerging from solely local interaction. The theoretical paradigm for the description of the transition to a synchronized state of many oscillating units is the Kuramoto model (Kuramoto, 1975; Acebron et al., 2005), a model for the dynamics of a large set of coupled oscillators. In the Kuramoto model the main parameters are the oscillators diversity, that acts as a source of disorder, and the interaction strength (coupling) that pushes the oscillators to synchronize together. Importantly enough, although in presence of disorder, the model is deterministic. The Kuramoto model is effective in many contexts and disciplines including biological systems, as it elucidated the core mechanisms of various biological phenomena, ranging from the rhythmic flashing of firefly congregations (Ermentrout, 1991) to the coordinate firing of neurons (Breakspear, 2010) or cardiac pacemaker cells (Osaka, 2017), corroborating what Winfree envisioned as the *geometry of biological time* (Winfree, 1991). Recently, (Scirè and Annovazzi-Lodi, 2017) a theoretical work, inspired by the Kuramoto model, introduced a deterministic phase transition with intrinsic self-organization properties, able to produce adaptive spatiotemporal patterns.

In this work, we enlighten a collective process (a synchronization process) due to which an ensemble of polar phase oscillators, free to move in a two-dimensional space, builds complex and self-regulating networks. Those oscillators build cooperative dynamic superstructures, that stretch over length scales much greater than the oscillators interaction length. Such networks compartmentalize the two-dimensional space with no a priori constraints, by means of porous transport walls. We therefore argue that our system displays generic life-like properties as self-organization and emergence, and structural ones like compartmentalization and transport walls.

The Manuscript is organized as follows, section 2 is devoted to the introduction and discussion of the model, section 3 shows the results for different punctual values of the control parameter, and contains a subsection named *Statistics*, where the averaged collective parameters are numerically evaluated by sweeping the control parameter. Finally, Section 4 is devoted to summarizing the manuscript and discussing the results.

## 2 The model

As any typical Artificial Life model (Maes, 1993; Bedau, 2003) our model is *bottom-up*, thus implemented as low-level agents that simultaneously interact with each other, and the dynamics of which is based on information about, and affects, only their own local environment. It relies on a recently introduced model (Scirè and Annovazzi-Lodi, 2017), that described the spatiotemporal dynamics of many coupled polar phase oscillators (the low-level agents) free to move in a two-dimensional space. The polar phase oscillators are abstract mathematical objects consisting of two complementary kinds of oscillators (poles) arbitrarily labelled as *circles* and *squares*. The circle and square poles obey an interaction law that makes them attract each other out-of-phase or mutually repel in-phase, as sketched in Fig. 1 (see (Scirè and Annovazzi-Lodi, 2017) for more details). Such poles are abstract agents, not aiming at modelling any physical object. However, the interaction scheme reminds the proton-electron dipole with spin interaction, the well-known Pauli exclusion principle, which shapes most of the natural chemistry. In (Scirè and Annovazzi-Lodi, 2017) we had shown how such system, despite its relative simplicity, retains high complexity and self-organizational properties by attributing random natural frequencies to the oscillators, and using the standard deviation of the statistical distribution as the control parameter of the analysis. Static "crystals" and dynamic "molecules" emerged and were discussed, without exhausting the possibilities of a surprisingly rich dynamics. Differently from (Scirè and Annovazzi-Lodi, 2017), here a fixed detuning is given between the circle and the square poles, leading to substantial new results. Such detuning is called $\Delta_H$ in the following, and it will be the control parameter of the analysis.

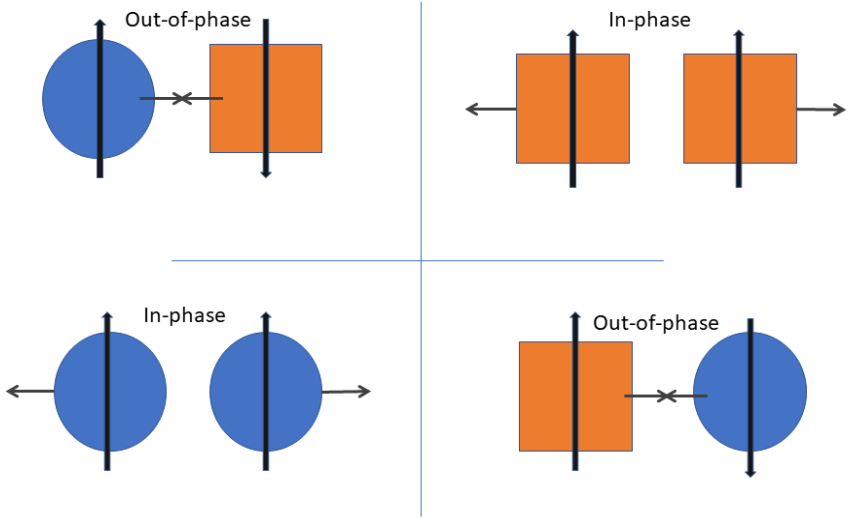

**Fig.1** Sketch of the dipolar interaction scheme, poles of different kind attract each other and their phases tent to be different by $\pi$ (out of phase), poles of the same kind repel each other and their phases tent to be equal (in-phase)

The equations of motion, modified from (Scirè and Annovazzi-Lodi, 2017) according to the above assumptions, for $N$ poles read

$$\dot{x}_i = \sum_{j=1}^{N} \nabla_i \, W(|x_i - x_j|) \cos(\varphi_i - \varphi_j), \qquad (1)$$

$$\dot{\varphi}_i = \gamma_i \Delta_H + \sum_{j=1}^{N} \gamma_i \gamma_j \, W(|x_i - x_j|) \sin(\varphi_i - \varphi_j), \qquad (2)$$

where $\nabla_i$ means differentiation respect to $x_i$. The potential wells (that make the interactions *local* with a characteristic length L=1) are chosen as exponential wells

$$W(|x_i - x_j|) = -e^{-|x_i - x_j|^2}, \qquad (3)$$

where $x_{i,j}$ are the spatial coordinates and $\varphi_{i,j}$ the local phases of the $i,j$-th pole with $i,j = 1,... N$, and $|x_i - x_j|$ is the Euclidean distance in space. The coefficients $\gamma_i$ express the pole polarity, i. e. $\gamma_i = 1$ if the pole is a square (*s-pole*) or $\gamma_i = -1$ if a circle (*c-pole*). Mathematically, the model (1)-(2) is a dissipative non-linear dynamical system, that can display chaotic attractors already with few elements (Scirè and Annovazzi-Lodi, 2017). The effect of $\Delta_H$ in Eq. (2) is to split the natural frequencies of *c* and *s*-poles and to represent a forcing term for the differential equations. In the following section, numerical results for the dynamics of many oscillators are shown, for different values of $\Delta_H$.

## 3 Results

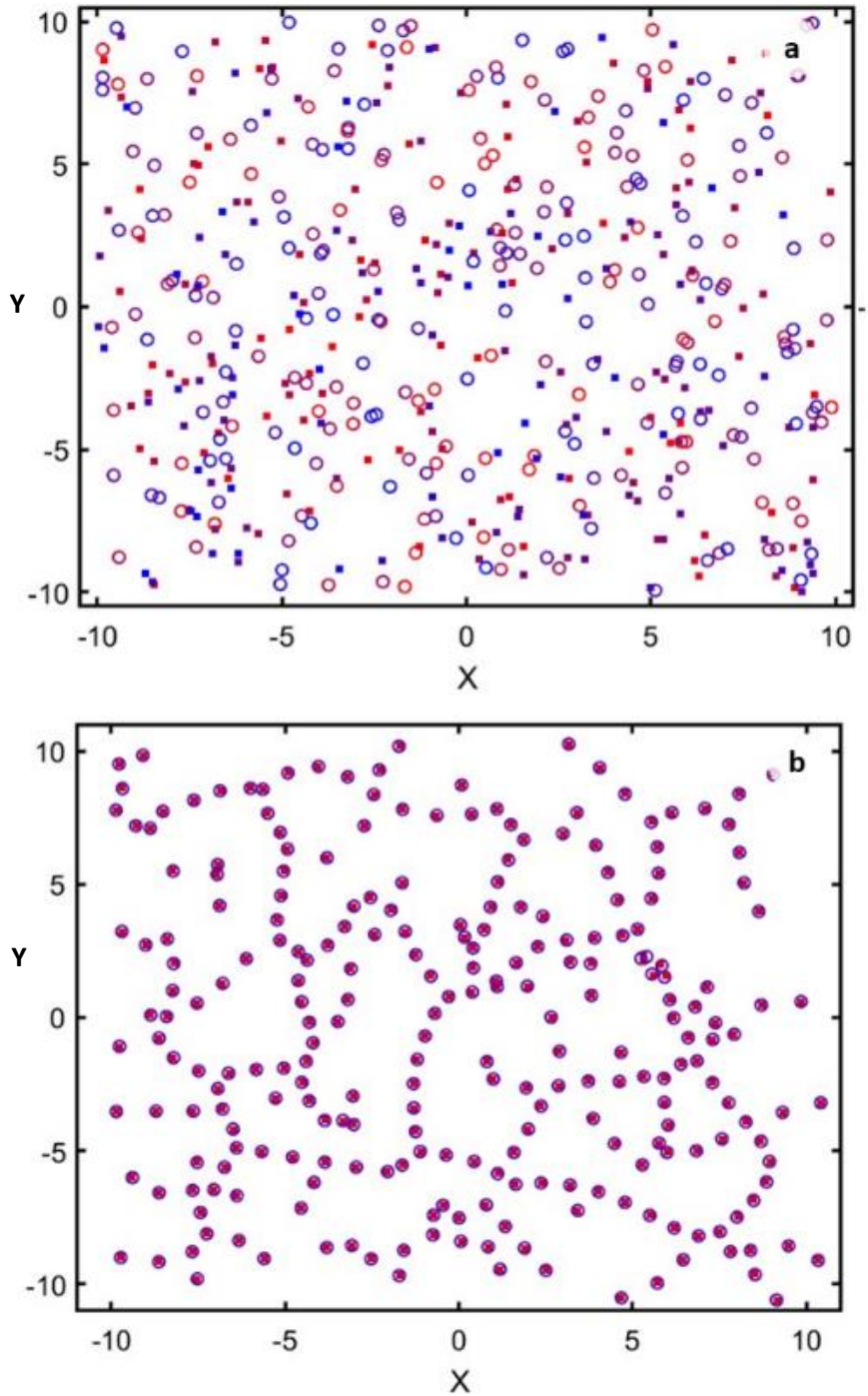

**Fig.2** a) Initial conditions drawn from a uniform distribution in a 20x20 square. b) Regime distribution after $2 \times 10^5$ integration steps simulation of Eqs. (1)-(2) for 500 poles (N=500). $N_{circles}$=250, $N_{squares}$=250. $\Delta_H = 0$

As a result of numerical simulations of Eqs. (1) - (2) for $\Delta_H = 0$ and N = 500 (corresponding to 250 c-poles and 250 s-poles), after a transient starting from random initial conditions (Fig.2a), a stable phase-locked global pattern is formed in the (X,Y) plane (Fig.2b). The oscillators end up located along chains characterized by some spatial regularity made by the superposition of a c and a s-poles, that attract each other out-of-phase and relax to occupy the same position (zero distance), i.e. any c-pole attains the same phase value, and any s-pole attains the same phase value, and the difference of those two values is $\pi$. Fig. 2 shows (a) the initial conditions, i.e. random positions and phases, and (b) the final pattern of overlapped dipoles. For each oscillator, the position is represented by spatial coordinates in a (X,Y) plane, whereas the phases $\varphi_{i,j}$ – considered in [0, $2\pi$] – are encoded into the *color* of the respective units by a standard colormap. From a visual point of view, c-poles are portraited by thin empty circles, whereas s-poles are portraited by smaller thick squares, so that when they overlap both markers are still visible. Concerning the spatial variables, initial random conditions are taken in a squared area such that at least one interaction is guaranteed in average, so that neither the space is too busy, and the pattern cannot develop, nor the units are too sparse, and they do not "see" each other. Initial phases are randomly distributed with a uniform distribution in [0, $2\pi$]. The movie S1 shows the progressive formation of the static regime pattern of Fig. 2b, from random starting conditions (Fig. 2a).

The degree of synchronization (phase locking) in a phase oscillator ensemble is conventionally quantified by the Kuramoto complex order parameter, which, slightly modified respect to the original formula (Kuramoto, 1975) due to the presence of the coefficients $\gamma_k$, reads

$$\rho e^{i\theta} = \frac{1}{N}\sum_{k=1}^{N}\gamma_k e^{i\varphi_k} \qquad . \qquad (4)$$

The absolute value $\rho$ measures the oscillators *global degree of entrainment*, it is bounded between 0 and 1, where 1 means total coherence (global phase locking) and 0 means total disorder (disordered phases, unlocking). The phase $\vartheta$ is the *global phase* of the whole ensemble, so the *global frequency $d\vartheta/dt$* can quantify collective pulsations, with regular ($d\vartheta/dt \sim$ constant) or chaotic dynamics ($d\vartheta/dt$ fluctuating), or collective excitability (temporal spikes in $d\vartheta/dt$).
Another useful global parameter, able to detect the kinetic motion in space, is the total kinetic energy

$$T = \sum_{k=1}^{N} v_k^2 \qquad , \qquad (5)$$

where $v_k = \dot{x}_k$ are the spatial velocities.
The order parameter (4) and the total kinetic energy (5) concerning the $\Delta_H = 0$ simulation above mentioned are shown in Fig. 3.

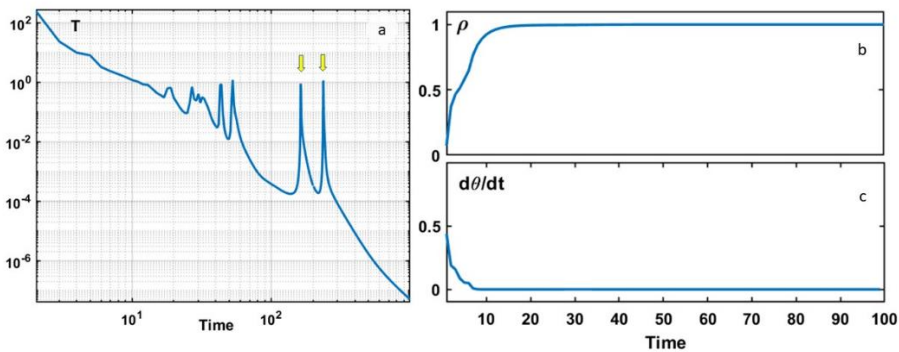

**Fig. 3** (a) Kinetic energy T Vs time (note the logarithmic scales to enhance the visibility of the transients). (b) Global degree of entrainment $\rho$ Vs time (c). Global frequency $d\vartheta/dt$ Vs time. $\Delta_H = 0$

Fig. 3 shows that, after a transient, the ensemble displays a static pattern (T $\rightarrow$0 in Fig.3a) and full static synchrony (phase-locking) with $\rho\rightarrow$1 (Fig.3b), and the global frequency $d\vartheta/dt\rightarrow$0 (Fig.3c).

During the transient the ensemble shows that the inclusion of the last elements in the pattern (blue arrows in movie S1) takes place by means of a collective involvement of the neighboring oscillators, causing peaks in the total kinetic energy (yellow arrows in Fig.3a). This is a sign that the formed pattern is a connected tissue, able to act collectively.

Increasing the detuning $\Delta_H$ a highly organized collective dynamics takes place. The movie S2 shows the dynamics starting from random conditions as in the previous case, but for $\Delta_H = 0.1$. The result is the formation of a spatiotemporal dynamic network of entrained currents, a synchronized flux of kinetic and phase dynamics, that shapes compartments with transport walls. Indeed, the movie S2 shows the formation of a dynamic network made of vesicles supporting a counterpropagating flux of *c* versus *s*-poles, continuously reshaping and reorganizing itself. The network represents a dynamic entrainment of a highly intricate nature, emergent, self-organized and – at best of our knowledge – never reported before. Due to the high nonlinearity of the interaction functions, small changes in the initial conditions leads to markedly different networks. The networks themselves are indeed complex chaotic attractors. Moreover, those networks establish for values of detuning $\Delta_H$ that would not give raise to phase current in a single cs-dipole, i.e. well below the excitable threshold of a single cs-dipole, see (Scirè and Annovazzi-Lodi, 2017). It is an emergent, collective, and cooperative effect.

As a static picture of the dynamics, an occupation matrix is calculated from above mentioned simulated data (see Fig. 4). The occupation matrix represents how many times a particle was present in a point in space during the simulation.

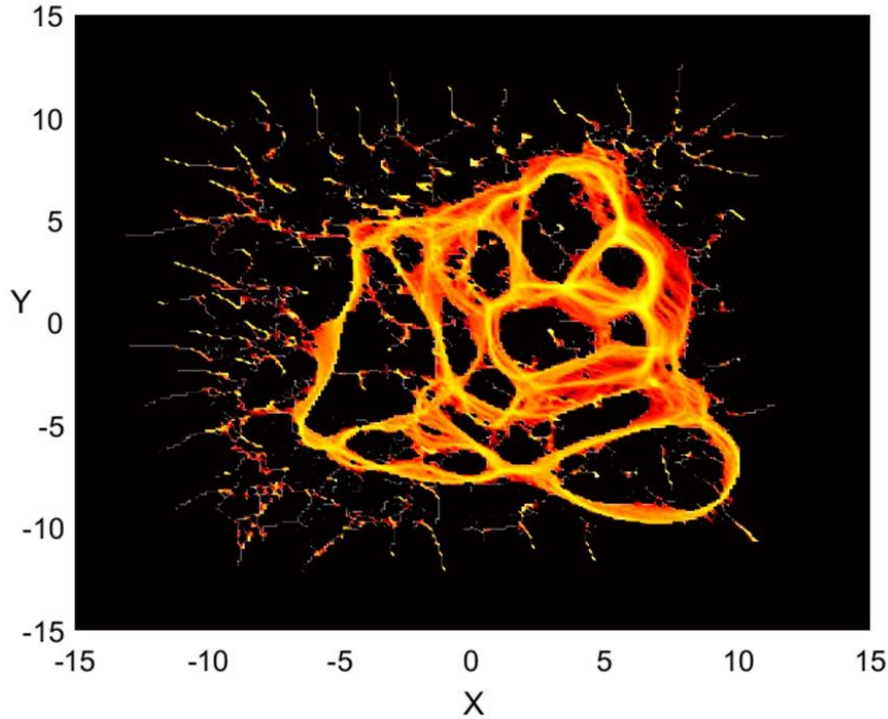

**Fig. 4** Occupation Map for $\Delta_H = 0.1$. Initial conditions drawn from a uniform random distribution in a 20x20 square and random phases in [0, 2$\pi$]. The picture concerns a simulation of Eqs. (1)-(2) for $10^7$ integration steps, sampling the dynamics each $10^3$ simulation steps, for 500 polar oscillators. $N_{circles}$=250, $N_{squares}$=250

Fig. 4 gives a static picture of the network and shows the persistence of vesiculation. For the same data, Fig. 5 shows: (a) the time dependent kinetic energy T(t) (b) the time evolution of the global entrainment $\rho(t)$, and (c) the global frequency $d\vartheta/dt$.

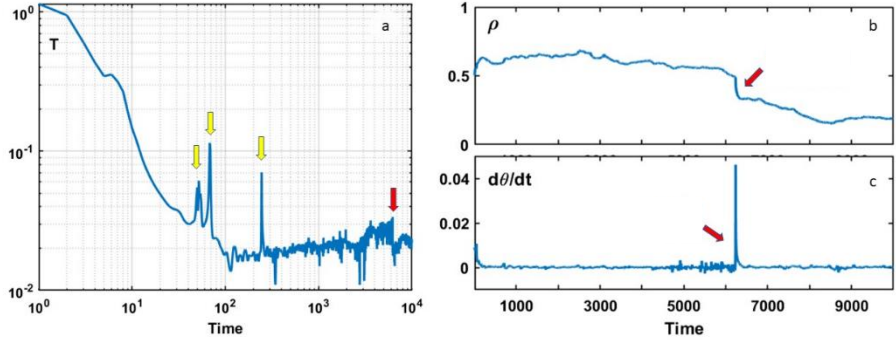

**Fig. 5** a) Kinetic energy T vs time. Order parameters: b) global entrainment $\rho$ and c) global frequency $d\vartheta/dt$ versus time, for $\Delta_H = 0.1$

Analyzing the movie S2 together with Fig.4 and Fig.5, we have devised three processes in the network evolution.

1) *Network Formation* (roughly $0 < t < 1000$). Local patterns, vesicles and chains of oscillators emerge from the initial disorder in different parts of the plane. They merge through collective events that cause peaks in the kinetic energy T(t) (yellow arrows in Fig.5a), until one single network is formed (roughly $800 < t < 1000$) by collective transport walls.

2) *Inclusion/exclusion of material from the environment*: (roughly $1000 < t < 6400$). An example is found at time $\approx 3700$, when a s-pole is included from the environment in the current flux, as shown by the movie S2 in the lower left corner and straight afterwards a dipole is expelled in the upper right part of the network (yellow arrows in movie S2). The transport walls appear therefore to be porous respect to the environment, while the network *self-regulates* the circulating material.

3) *Network Uniformation*: (roughly $t > 6400$). The movie S2 shows that the network is the results of an adaptive merging of initially separated patterns, including small networks that possess different flux velocities. During the process of forming a unique network the flux velocities of the distinct vesicles undergo changes, to be compatible with the whole structure. This happens by means of smooth changes or abrupt events as the one that takes place close after time = 6400, when two vesicles merge their flux (red arrow in movie S2) and the total kinetic energy suddenly lowers (red arrow in Fig.5a). Indeed, the merging of the two vesicles appears to be functional to the overall flux as the system globally slows down and spontaneously uniformizes the velocities in the network, that can be now sustained while using less kinetic energy. Such event is as well signalized by both a time peak in the global frequency $d\vartheta/dt$ (called a *collective firing* in (Tessone et al., 2007) – see the red arrow in Fig. 5c) and by a sudden lowering (red arrow in Fig 5b) in the entrainment $\rho(t)$. This is a typical behavior associated to the collective firing in coupled oscillating systems, as already reported in (Tessone et al., 2007) "*We […] show that the mechanism for collective firing is generic: it arises from degradation of entrainment.*"

Processes 2 and 3 may overlap and repeat in time and appear to be functional to the network persistency and self-regulation, at least during the investigated simulation time.

In order to evaluate the robustness of the networks, we have performed several numerical simulations with different ensembles, always retaining global "neutrality", i.e. the same number of c and s-poles. We have observed that networks need a consistent number of oscillators to emerge, of the order of N ~ 200. Moreover, in order to investigate the long-term behavior, we have performed long simulations. The movie S3 reports of a long simulation for N = 400 and $\Delta_H$ = 0.15, where the initial transient has been removed for brevity, showing the persistence of the network for over $10^9$ simulation steps, in compresence with smaller static patterns.

Increasing further the detuning $\Delta_H$, intermittency between the formation of dynamic networks and low dimensional patterns takes place. As an example, for $\Delta_H$ = 0.2 the network initially stabilizes, but after a while it breaks up in smaller patterns that prevent long term stable transport walls, as shown in movie S4. The occupation matrix for $\Delta_H$ = 0.2 is shown in Fig. 6, and it is shaped by the irregular compresence of vesicles, chains, small dynamic patterns and single poles.

The order parameters for $\Delta_H$ = 0.2 exhibit (Fig.7) high kinetic activity and coherence degradation after a transient where coherent vesicles were present, and the activity was relatively stable (roughly from t = 200 to t = 2000). Increasing further the detuning $\Delta_H$ the dynamics becomes progressively more erratic, and no trace of order is finally left. Remarkably, the system does not include any noise or disorder but the starting conditions, the progressively disordering observed scenario is a consequence of a non-linear dynamics.

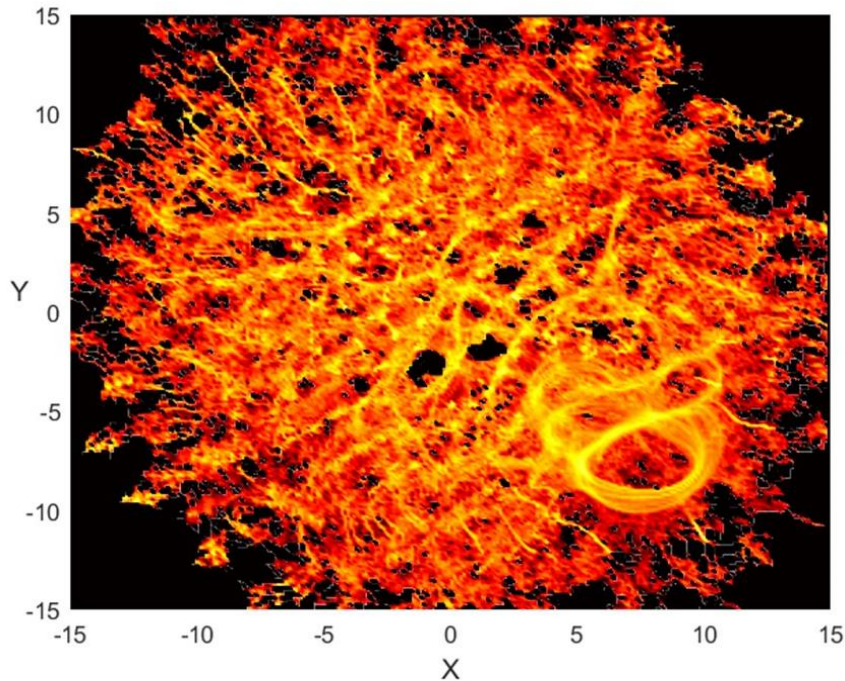

**Fig. 6** Occupation Matrix for $\Delta_H$ =0.2. Initial conditions drawn from a uniform distribution in a 20x20 square and random phases in [0, 2$\pi$], simulation of $10^7$ integration steps simulation for 500 poles. $N_{circles}$=250, $N_{squares}$=250

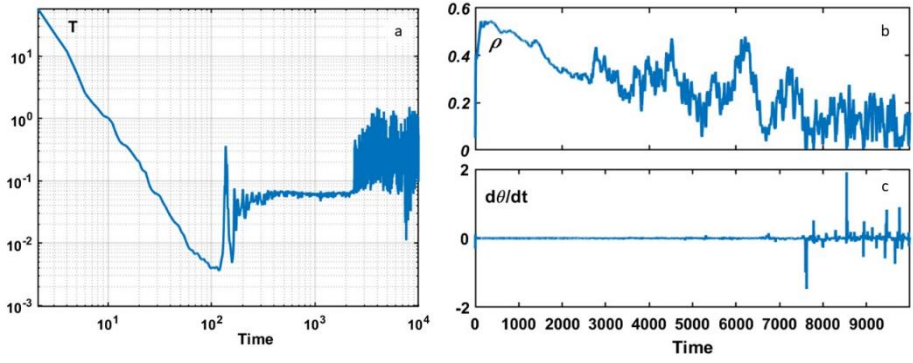

**Fig. 7** a) Kinetic energy T vs time. Order parameter: b) global entrainment $\rho$ and c) global frequency $d\vartheta/dt$ versus time, for the same simulation as in Fig.6

To evaluate the robustness of the dynamic network solution, further numerical simulations were performed in presence of small additive white noise or diversity in the natural frequencies. Preliminary results showed the persistence of the above depicted scenario in both circumstances.

*Statistics*

This subsection illustrates the behavior of the time averaged global parameters versus the detuning $\Delta_H$. The obtained results for a given ensemble are summarized in Fig. 8. Fig. 8A shows the time averaged total kinetic energy $<T>$ and highlights different regimes regions. Figure 8B shows the time averaged global entrainment $<\rho>$, and Fig. 8C shows the time averaged collective frequency $<d\vartheta/dt>$, all these numerically calculated versus the control parameter $\Delta_H$. For $\Delta_H \sim 0$, synchronized static patterns are formed, the kinetic energy vanishes, the global entrainment is maximum close to 1, and the global frequency is stationary close to zero.

Increasing $\Delta_H$, for $\Delta_H \sim 0.05$, the system shows the onset of coherent currents, shaping dynamic networks with transport walls that persists up to $\Delta_H \sim 0.2$. Here the kinetic energy is non-vanishing in order to dissipate the energy fed to the system by the detuning, the degree of entrainment is not vanishing ($<\rho> \sim 0.5$) and the global frequency ($<d\vartheta/dt> \sim 0$) is stable, so there is a significative degree of global coherence.

Numerical simulations showed that vesicles (the compartments of the networks) become smaller increasing the detuning for a given ensemble, but they become also intermittent and unstable. For $\Delta_H \sim 0.2$, the onset of an intermittent regime that includes disordered dynamic patterns compromises the networks stability. The average kinetic energy increases because more energy is now fed to the system. In general, the emerging patterns are dissipative structures that serve to dissipate the energy fed to the system by the detuning $\Delta_H$. When the detuning is low, the slow dynamics of the big networks is effective as a dissipative structure, but when the detuning is increased, smaller dynamic structures, able to move faster, are preferred for that purpose. Panels B and C in Fig. 5 show that coherence is progressively lost when $\Delta_H > 0.2$, because local (instead of collective) dynamics prevails.

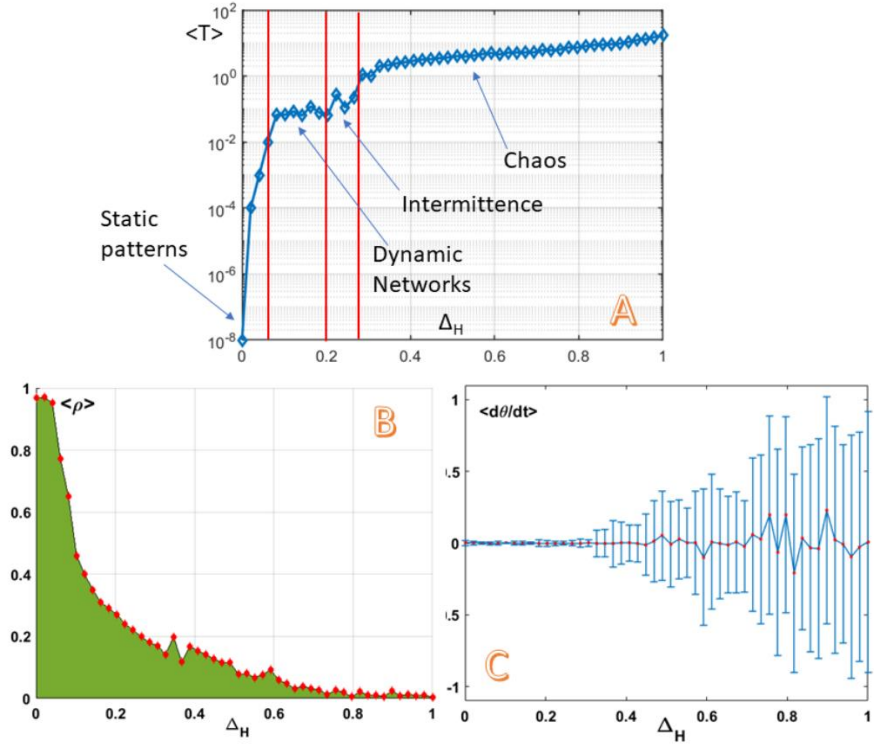

**Fig. 8** *Statistics*. A: Time averaged Kinetic energy vs $\Delta_H$, with indications concerning the different regimes. B: Time averaged collective entrainment $\rho$ vs $\Delta_H$. C: Time averaged (with standard deviation) global frequency Vs $\Delta_H$. For each simulation: Initial conditions drawn from a uniform distribution in a 20x20 square, simulation of $2\times10^6$ integration steps for N = 500. $N_{circles}$=250, $N_{squares}$=250

The initial oscillators density also revealed to be of some importance. If the oscillators start too tight (many oscillators within the interaction length) the space is too busy, the networks are compressed and may fail to develop. If the initial density is too low (much less than one oscillator in average within the interaction length) the ensemble disaggregates in subdomains that do not interact each other. As a rule of thumb, a good value (without being critical) for the initial density $d = N/l^2$, where $l$ is the square side of the initial conditions area, is $d \sim 1$.

Open questions concern the asymptotic stability of the networks: whether they keep indefinitely evolving or finally attain a fixed configuration, and whether their lifetime is infinite or not. Those issues will be addressed in future investigations.

## 4 Conclusions

We have reported of an ensemble of polar phase oscillators free to move in a two-dimensional space, that in certain conditions build coherent dynamic networks. Those networks are emergent and self-regulating complex superstructures, resulting from of a cooperative behavior involving many units over length scales much greater than the interaction length, compartmentalizing the two-dimensional space with no a priori constraints. This kind of behavior has – at best of our knowledge – never been reported before in soft artificial life systems, neither in theoretical soft matter, nor in dynamical systems in general. The analysis was numerically carried out as a function of a control parameter, showing static pattern formation, the emergence of persistent and intermittent dynamic networks, and irregular dynamics, respectively, for increasing values of the control parameter. Such complex scenario is solely due to the non-linear dynamics of the system, where the starting conditions are the only source of disorder included.

We have drawn a numerical statistical analysis versus the control parameter, to have a glance over the whole scenario, identifying the range of existence of the different regimes for a given ensemble. The robustness issue against noise and diversity has been checked with good preliminary results.

In conclusion, we have argued that our system displays emergence, complexity, self-organization, spontaneous compartmentalization and self-regulation, only due to a non-linear many-body dynamics. This model is expected to contribute to the field of artificial life and, in general, it gives a new portrait of the organizational processes that govern the emergence of adaptive structures from locally interacting units.

## Acknowledgements

A.S. Acknowledges Giuseppe Aromataris (University of Pavia, Pavia - Italy) and Emilio Hernandez-Garcia (Instituto de Fisica Interdisciplinar y Sistemas Complejos, Palma de Mallorca - Spain) for fruitful conversations.

## Supporting information captions

### S1 movie
Spatio-temporal dynamics resulting from a numerical simulation of Eqs. (1)-(2). N = 500 (250 c-poles and 250 s-poles). $\Delta_H = 0$.

### S2 movie
Spatio-temporal dynamics resulting from a numerical simulation of Eqs. (1)-(2). N = 500 (250 c-poles and 250 s-poles). $\Delta_H = 0.1$. A time counter has been added to the movie in order to better connect the spatio-temporal dynamics to Fig.5.

### S3 movie
Spatio-temporal dynamics resulting from a numerical simulation of Eqs. (1)-(2). N = 400 (200 c-poles and 200 s-poles). $\Delta_H = 0.15$.

### S4 movie
Spatio-temporal dynamics resulting from a numerical simulation of Eqs. (1)-(2). N=500 (250 c-poles and 250 s-poles). $\Delta_H = 0.2$.